\documentclass[runningheads,english]{llncs}
\usepackage{graphicx}
\usepackage{pgfplots}
\usepackage{makeidx}  
\usepackage{babel}
\usepackage{array}
\usepackage{url}
\usepackage{bm}
\usepackage{multirow}
\usepackage{amstext}
\usepackage{amssymb}
\usepackage{amsmath}
\usepackage{pdflscape}
\usepackage[unicode=true,pdfusetitle,bookmarks=true,bookmarksnumbered=false,bookmarksopen=false,breaklinks=false,pdfborder={0 0 1},backref=false,colorlinks=false]{hyperref}
\usetikzlibrary{arrows,decorations.pathmorphing,backgrounds,positioning,fit,petri,matrix}

\newcommand{\norm}[1]{\left\lVert#1\right\rVert}
\def\versionLong{1}
\begin{document}
	\title{3d-SMRnet: Achieving a new quality of MPI system matrix recovery by deep learning}

	\titlerunning{3d-SMRnet: MPI system matrix recovery by deep learning}
	
	\if\versionLong0
	    \author{Anonymous}
	    \institute{Famous Institute, State, Country\\
		\email{name@inst.tld, https://www.inst.tld/group}}
	\else
	
	\author{Ivo M. Baltruschat\inst{1,2,3}, Patryk Szwargulski\inst{1,2,3}, Florian Griese\inst{1,2}, Mirco~Grosser\inst{1,2}, Rene~Werner\inst{3,4}, Tobias Knopp\inst{1,2}}
	\institute{Section for Biomedical Imaging, University Medical Center Hamburg-Eppendorf
		\and Institute for Biomedical Imaging, Hamburg University of Technology
		\and DAISYLabs, Forschungszentrum Medizintechnik Hamburg, Germany
		\and Department of Computational Neuroscience, University Medical Center Hamburg-Eppendorf\\
		\email{i.baltruschat@uke.de, \url{https://www.tuhh.de/ibi}}}
	\fi

	\if\versionLong0
	    \authorrunning{Anonymous}
	\else
	    \authorrunning{I. M. Baltruschat et al.}
	\fi 
	
	\maketitle              
	
	\begin{abstract}
		Magnetic particle imaging (MPI) data is commonly reconstructed using a system matrix acquired in a time-consuming calibration measurement. The calibration approach has the important advantage over model-based reconstruction that it takes the complex particle physics as well as system imperfections into account. This benefit comes for the cost that the system matrix needs to be re-calibrated whenever the scan parameters, particle types or even the particle environment (e.g. viscosity or temperature) changes. One route for reducing the calibration time is the sampling of the system matrix at a subset of the spatial positions of the intended field-of-view and employing system matrix recovery. Recent approaches used compressed sensing (CS) and achieved subsampling factors up to 28 that still allowed reconstructing MPI images of sufficient quality. In this work, we propose a novel framework with a 3d-System Matrix Recovery Network and demonstrate it to recover a 3d system matrix with a subsampling factor of 64 in less than one minute and to outperform CS in terms of system matrix quality, reconstructed image quality, and processing time. The advantage of our method is demonstrated by reconstructing open access MPI datasets. The model is further shown to be capable of inferring system matrices for different particle types.
		
	\end{abstract}

	\section{Introduction}
	
	Magnetic particle imaging (MPI) is a young tomographic imaging technique that quantitatively images magnetic nanoparticles with a high spatio-temporal resolution and is ideally suited for vascular and targeted imaging \cite{Gleich2005TomographicIU}. 
	One common way to reconstruct MPI data is the system matrix (SM)-based reconstruction \cite{Knopp2010PhysMedBio}. It requires a complex-valued SM, which is currently determined in a time-consuming calibration measurement. A delta sample is moved through the field-of-view (FOV) using a robot and the system response is measured in a calibration process. The number of voxels encoded in the SM directly determines the image size but also the scan time. The acquisition of a $37\times 37\times 37$ voxel SM takes about 32 hours,  compared to an $9\times 9\times 9$ SM, which takes about 37 minutes. Therefore, a compromise between image size and scan time is usually made.
	While in principle the calibration of the SM needs to be done only once, the resulting SM is only valid for a very specific set of scan parameters. When changing scan parameters such as the size or the position of the FOV, the SM calibration needs to be redone.
	Furthermore, the SM highly depends on the type of the particles and their binding state, viscosity and even temperature. This makes it almost impossible to record high-resolution 32-hour system matrices for each combination of scan parameters and particle settings.
	
	The first work that investigated calibration time reduction for MPI applied a compressed sensing (CS) to recover a subsampled 2d SM with 10-fold subsampling \cite{knopp2013sparse}. The basic idea is to exploit the fact that the MPI system matrix rows (i.e. frequency components) are consisting of wave-like patterns with a certain oscillation degree. By applying a discrete cosine transform (DCT), each matrix row can be sparsified such that CS with an $L_1$ prior can be applied. Motivated by the success, recent work proposed a CS approach by combining CS with a method that exploits symmetries in the SM. For up to 28-fold subsampling, sufficient image quality after reconstruction was archived \cite{weber2015reconstruction}. 
	
	While the CS approach for SM recovery is promising, it still leaves room for improvement since the sparsification using the DCT is not perfect and row specific. Furthermore, the CS approach currently cannot take prior knowledge from existing high resolution (HR) system matrix measurements into account. In the present work, we will, for the first time, investigate if deep learning (DL) techniques can be used to improve the SM recovery problem in MPI. DL-based super-resolution techniques have been demonstrated to be superior in the up-scaling of images in computer computer vision \cite{Dong2014,Tai2017} and recently also for 3d medical image up-scaling \cite{Wang2016MRI,Chen2018BrainMS,Chen2018}. While Super-Resolution Convolutions Networks (SRCNNs) and Super-Resolution Generative Adversarial Networks (SRGANs) are mostly used directly in the image domain, it is inherently difficult to restore texture and structural details. SRGANs have proven to be successful in CV and medical image processing in modeling visually more appealing images than SRCNNs; SRCNNs typically tend to blur the SR image as a whole. Yet, this property may potentially be beneficial if SR is applied \emph{prior} to image reconstruction -- like in the current case for MPI SM recovery. 
	
\begin{figure}[htb!]
        \vspace{-1em}
		\includegraphics[width=\textwidth]{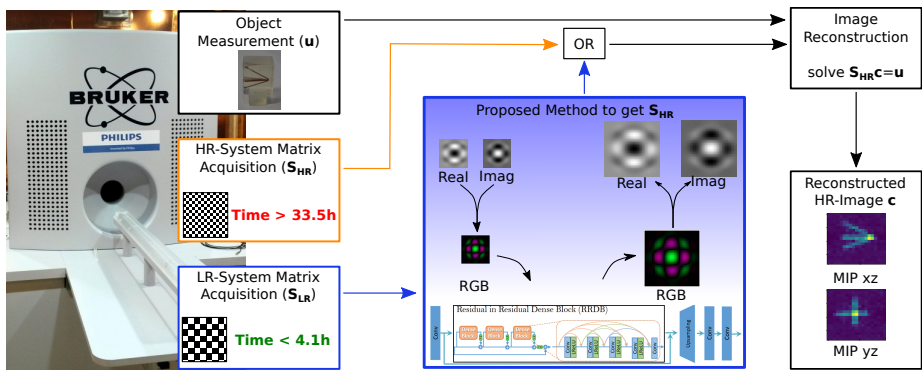}
		\caption{Overview of the data flow when employing our proposed 3d-SMRnet (\textcolor{blue}{blue path}). Instead of measuring a HR SM and using it for reconstruction (\textcolor{orange}{orange path}), only a LR SM (lower left) is measured and the HR SM is retrieved by applying our proposed method to each frequency component of the LR SM. The recovered HR SM can be used for reconstruction (upper right). 
		} 
		\label{fig:overview}
		\vspace{-3em}
	\end{figure}

	To evaluate the potential of SRCNN-based MPI SM recovery, we present a novel framework that comprises three central steps (see proposed method branch in Fig. \ref{fig:overview}). First, we acquire a low resolution (LR) SM on a specific sampling grid. Secondly, we encode each complex number of the 3d-SM to RGB vectors, allowing us to leverage SRCNNs from CV or medical image processing. Thirdly, we employ a SRCNN, which we call 3d System Matrix Recovery network (3d-SMRnet), to recover a high resolution (HR) SM by adapting the model to work on 3d RGB input data and employing it to each frequency component of the SM. Finally, we decode each RGB vector of the high resolution SM back to a complex number and use this newly recovered SM to reconstruct a high resolution image.
	
	We evaluate our method in Sec. \ref{sec:experiment} on the Open MPI Data and will show that our framework reaches superior performance in image quality, SM quality, and processing time compared to the current state-of-the-art.  
	All afore-mentioned aspects (introduction of 3d-SMRnet; conversion of MPI raw data to RGB format; comparison of 3d-SMRnet to compressed sensing) are novel contributions.
	
	\section{Methods}
	In MPI, the relation between the particle concentration $c(\cdot)$ and the Fourier coefficients $\hat{u}_k$ of the induced voltage $u(t)$ at frequency $f_{k}$ can be described by the linear integral equation $\hat{u}_k = \int_\Omega \hat{s}_k(\mathbf{r})c(\mathbf{r})\text{d}^3r$ where $\mathbf{r}$ is the spatial position, $k$ is the frequency index, and $\hat{s}_k(\cdot)$ denotes the system function. By sampling the FOV at $N$ positions $\mathbf{r}_n, n=0,\dots,N-1$, one obtains a linear system of equations written in matrix-vector form $\mathbf{Sc}=\hat{\mathbf{u}}$ where $\hat{\mathbf{u}}=(\hat{u}_{k})_{k=0}^{K-1} \in \mathbb{C}^K$ and $\mathbf{c}=(c(\mathbf{r}_{n}))_{n=0}^{N-1} \in \mathbb{R}^N$ are the measurement vector and the particle concentration vector, respectively. $K$ is the total number of frequency components.
	 The goal of this work is to recover a high resolution system matrix
	\vspace{-0.1cm}
	\begin{equation}
	    \mathbf{S}=(\hat{s}_k(\mathbf{r}_n))_{k=0,\dots,K-1;n=0,\dots,N-1} \in \mathbb{C}^{K \times N}
	    \vspace{-0.1cm}
	\end{equation}
	using a subset of the total number of sampling positions $N$. In our work, we treat the SM rows as independent images and train a network based on the entire set of rows of a measured HR SM. Then for SM recovery one measures a LR SM and infers the HR SM from the trained network.
	The method consists of three main steps, which are outlined in Fig.~\ref{fig:overview} and explained in detail below.
	
	\textbf{System Matrix Sampling:}
    Usually, the system matrix is acquired using a 3-axis linear robot to scan the FOV equidistantly on a predefined grid. In this way, both HR and LR system matrix can be acquired. 
    This is comparable to the generation of a LR SM by sampling every $\text{n}^{\text{th}}$ voxel from a HR SM. We employ the latter to get a LR and HR SM pair as training data for our 3d-SMRnet.
	
	\textbf{RGB Encoding and Decoding of System Matrix:}
	Before we feed the frequency components of the SM into our 3d-SMRnet, we transform each complex number $\hat{s}_k(\mathbf{r}_n)$ of the SM to a RGB color vector $s^{\text{RGB}}_k(\mathbf{r}_n) \in \mathbb{R}^3$ in two steps. First, we use the hue-saturation-value (HSV) color model to represent the phase $\arg \hat{s}_k(\mathbf{r}_n)$ with hue following the color wheel. Therefore, we employ the transformation $T_{\text{HSV}}: \mathbb{C} \to \mathbb{R}^3$ with $T_{\text{HSV}}(\hat{s}_k(\mathbf{r}_n)) = (H, S, V) = (\arg \hat{s}_k(\mathbf{r}_n), 1, 1)$, where we omit S and V by setting them to 1. Secondly, we convert the HSV color vector $s^{\text{HSV}}_k(\mathbf{r}_n) = T_{\text{HSV}}(\hat{s}_k(\mathbf{r}_n))$ to $s^{\text{RGB}}_k(\mathbf{r}_n)$ with a standard HSV to RGB transformation \cite{ford1998colour}. 
	Finally, the amplitude $|\hat{s}_k(\mathbf{r}_n)|$ is coded by the intensity. Hence, we linearly scale the RGB color vector by the amplitude.
	
	For decoding, we recover the complex numbers by extracting the scaling factor of the RGB color vector for the amplitude. Afterwards, we normalize the RGB color vector by the amplitude and convert it to a HSV color vector. The phase is now the hue value.
	
	\textbf{3d-System Matrix Recovery Network: }
	Following \cite{Wang2019ESRGAN} and \cite{ledig2017photo}, we extend the SRCNN with Residual-In-Residual-Dense-Blocks (RRDBs) from 2d-RGB to 3d-RGB image processing. Our model contains two branches: image reconstruction and feature extraction. The feature extraction branch consists of $R$ stacked RRDBs (here $R = 9$). Each RRDB combines three dense connected blocks and four residual connections as illustrated in Fig.~\ref{fig:overview}. The dense connected blocks are built upon five convolutional layers. The image reconstruction branch generates the final up-scaled image by $U$ up-convolution blocks, followed by two convolutional layers. The up-convolution block contains a nearest-neighbor interpolation upsampling and a convolutional layer as proposed by \cite{odena2016deconvolution} to reduce checkerboard artifacts from deconvolution. In our model, all 2d convolutions are replaced by 3d convolutions. Hence, the 3d-SMRnet recovers a HR SM by employing it to each frequency component $K$ of a LR SM.
	
	\vspace{-1em}
	\section{Materials and Experiments}
	\label{sec:experiment}
	We apply our framework to the Open MPI Data\footnote{\url{https://magneticparticleimaging.github.io/OpenMPIData.jl/latest/}} dataset; It contains two HR system matrices, one for the particles Perimag $\mathbf{S}^{\text{Peri}}_{\text{HR}}$ and another for the particles Synomag-D $\mathbf{S}^{\text{Syno}}_{\text{HR}}$. Both are acquired using a $4~\text{}\mu$L delta sample with a concentration of 100 mmol/L and a grid size of $37\times37\times37$. Hence, $\mathbf{S}^{\text{Peri}}_{\text{HR}}$ and $\mathbf{S}^{\text{Syno}}_{\text{HR}}$ have the dimensions $37\times37\times37\times K$.  Furthermore, three different phantom measurements with Perimag are provided in the Open MPI Data: Shape Phantom, Resolution Phantom, and Concentration Phantom. In our experiment, we train our 3d-SMRnet on Synomag-D with frequency components of the subsampled $\mathbf{S}^{\text{Syno}}_{\text{LR}}$ as input against $\mathbf{S}^{\text{Syno}}_{\text{HR}}$ and test it on the subsampled $\mathbf{S}^{\text{Peri}}_{\text{LR}}$. This represents the interesting case where the SM for new particles is inferred from a network trained on an established particle system. In addition, this approach prevents overfitting of the data.
	
	We evaluate the recovered SM results in two steps. First, we compare all recovered system matrices with the ground truth $\mathbf{S}^{\text{Peri}}_{\text{HR}}$ by calculating the normalized root mean squared error (NRMSE) for each frequency component. Secondly, we reconstruct the measurements of the Open MPI shape, resolution and concentration phantoms with all recovered system matrices using the same standard regularization parameter ($\lambda=0.01, \text{iter}=3$).
	\textbf{Implementation Details and Training: }
	We implement three versions 3d-SMRnet$_{8\times}$, 3d-SMRnet$_{27\times}$, and 3d-SMRnet$_{64\times}$. For 3d-SMRnet$_{8\times}$ and 3d-SMRnet$_{27\times}$, we set $U=1$ and 2-times and 3-times upsampling, respectively. Furthermore, 3d-SMRnet$_{64\times}$ uses $U=2$ and 2-times upsampling to finally upsample 4-times.
	To generate a LR and HR SM pair for training, we zero-pad $\mathbf{S}^{\text{Syno}}_{\text{HR}}$ to $40\times40\times40$ with two rows and one row at the beginning and the end, respectively. Afterwards, we apply 8-fold, 27-fold, and 64-fold subsampling, resulting in $20\times20\times20$, $13\times13\times13$ and $10\times10\times10$ spatial dimensions for the input volume, respectively. After applying a threshold with a signal-to-noise ratio (SNR) of 3, $\mathbf{S}^{\text{Peri}}_{\text{HR}}$ and $\mathbf{S}^{\text{Syno}}_{\text{HR}}$ have $K=3175$ and $K=3929$ frequency components, respectively. We split $K$ of $\mathbf{S}^{\text{Syno}}_{\text{HR}}$ into 90\% training and 10\% validation data. 
	
	We use random $90^\circ$ rotations and random flipping as data augmentation. In total, we train for $2 \cdot 10^5$ iterations. Each iteration has a minibatch size of 20 for 3d-SMRnet$_{8\times}$ and 64 for 3d-SMRnet$_{27\times}$ and 3d-SMRnet$_{64\times}$. For optimization, we use ADAM with $\beta_1=0.9$, $\beta_2=0.999$, and without regularization. We trained with a learning rate of $10^{-5}$ for 3d-SMRnet$_{8\times}$ and $10^{-4}$ for 3d-SMRnet$_{27\times}$ and 3d-SMRnet$_{64\times}$. The learning rate is reduced by two every $4 \cdot 10^3$ iteration. As loss function, we employ the \textit{mean squared error} (MSE). Our models are implemented in PyTorch and trained on two Nvidia GTX 1080Ti GPUs. To support the reproduction of our results and further research, our framework and code are publicly available at 
	\if\versionLong0
	    \textit{https://github.com/''will be inserted upon acceptance``}.
	\else
	    \url{https://github.com/Ivo-B/3dSMRnet}.
	\fi 
	
	\textbf{Comparison to State-of-the-Art: }
	Compressed sensing exploits the fact that the SM becomes sparse when a discrete cosine transform (DCT) is applied to its rows. As shown in \cite{knopp2013sparse,donoho2006cs}, such signals can be recovered from an subsampled measurement by solving regularized least squares problems of the form
	\vspace{-0.15cm}
    \begin{equation}
        \underset{\mathbf{s}_k}{\operatorname{min}} \norm{      \bm{\Phi}\mathbf{s}_k }_1 \text{subject to } \mathbf{P}\mathbf{s}_k = \mathbf{y}_k.
        \label{eq:csprob}
        \vspace{-0.15cm}
    \end{equation}
	Here, $\mathbf{y}_k \in \mathbb{C}^M$ contains the values of the $k^{\text{th}}$ SM row at the measured points and $\mathbf{P}\in \mathbb{C}^{M\times N}$ is the corresponding sampling operator. Moreover, $\mathbf{s}_k\in \mathbb{C}^N$ is the SM row to be recovered and $\bm{\Phi}\in \mathbb{C}^{N\times N}$ denotes the DCT-II.
	
	Since CS requires an incoherent sampling, it cannot be applied directly to the regular sampled LR SMs used for our 3d-SMRnet. Instead, we use 3d Poisson disc patterns to subsample $\mathbf{S}^{\text{Peri}}_{HR}$ and obtain incoherent measurements with the same number of samples as used by the 3d-SMRnet. For every frequency component, we then normalize the measurement $\mathbf{y}_k$ and solve \eqref{eq:csprob} using the Split Bregman method. The solver parameters are chosen manually such that the average NRMSE for all frequency components is minimized.
	
	\section{Results and Discussion}
	
	\begin{figure}[htb!]
        \vspace{-1em}
		\includegraphics[width=\textwidth]{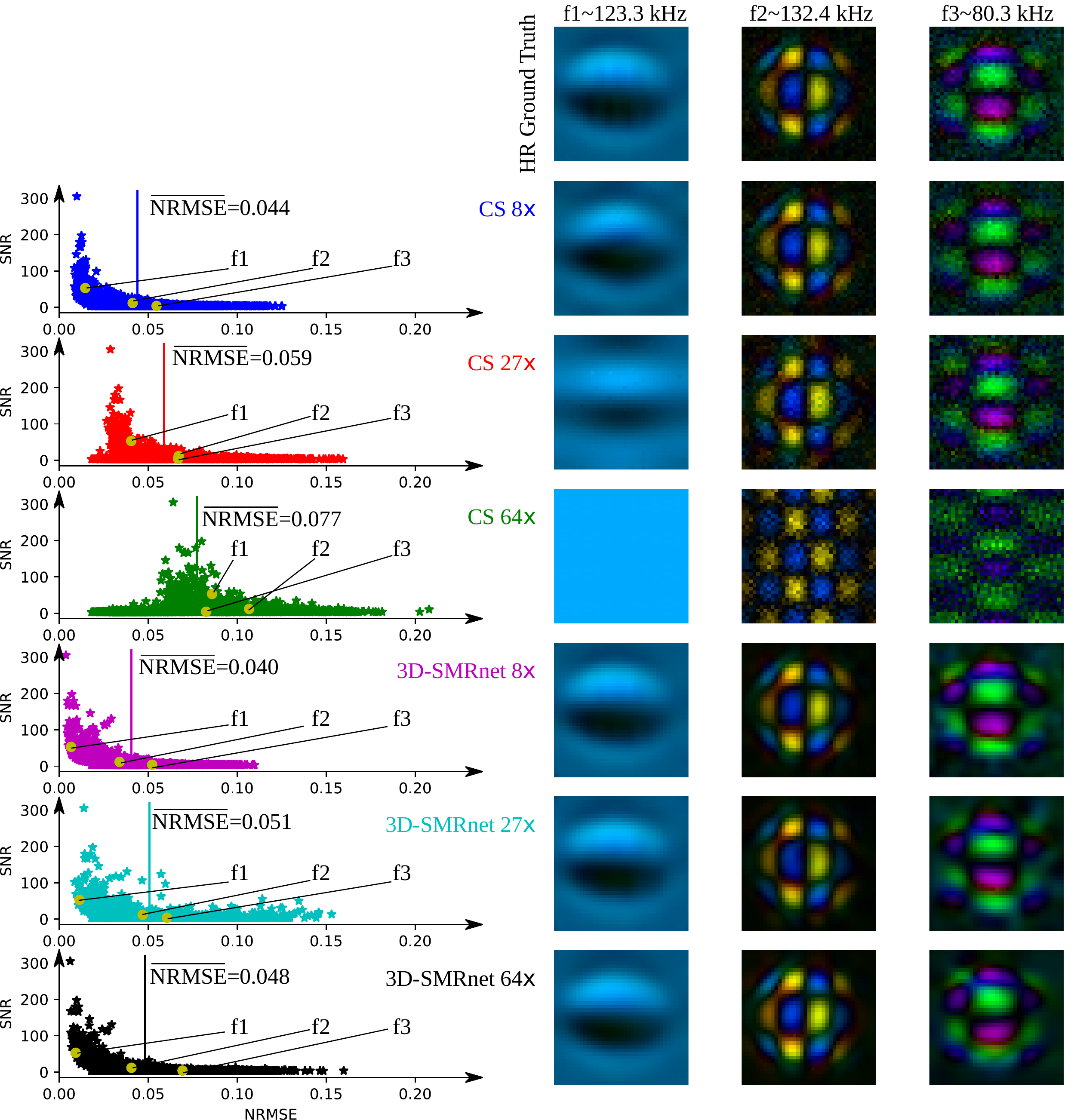}
		\caption{(left) SNR for each frequency component plotted against their NRMSE for recovered system matrices $\mathbf{S}_{\text{CS}}^{\text{8x}}$, $\mathbf{S}_{\text{CS}}^{\text{27x}}$, $\mathbf{S}_{\text{CS}}^{\text{64x}}$, $\mathbf{S}_{\text{3d-SMRnet}}^{\text{8x}}$, $\mathbf{S}_{\text{3d-SMRnet}}^{\text{ 27x}}$ and $\mathbf{S}_{\text{3d-SMRnet}}^{\text{64x}}$. (right) Representation of system matrix patterns of exemplary frequency components $f_1$,$f_2$ and $f_3$ for all recovered system matrices and ground truth. The frequency components are presented as RGB converted data} 
		\label{fig:SystemmatrixErrorNRMSE}
		\vspace{-1em}
	\end{figure}
	 Fig.~\ref{fig:SystemmatrixErrorNRMSE} (left) shows the NRMSE plot and Fig.~\ref{fig:SystemmatrixErrorNRMSE} (right) shows visualizations of recovery for three frequencies. The results show that all our 3d-SMRnets can correct the noisy characteristics in $\mathbf{S}^{\text{Peri}}_{\text{HR}}$ due to smoothing characteristics of training with a MSE loss function, whereas CS cannot. For all three reduction factors, the 3d-SMRnet has a lower mean NRMSE than CS: 0.040 vs 0.044, 0.048 vs 0.051, and 0.048 vs 0.077 for 8-fold, 27-fold, and 64-fold subsampling, respectively. While CS cannot sufficiently recover the SM for 64-fold subsampling, our 3d-SMRnet$_{64\times}$ still recovers the SM with a $37.66\%$ lower NRMSE. This low NRMSE is comparable to the results of CS with a 8-fold subsampling. Furthermore, our model 3d-SMRnet$_{64\times}$ is over $42$ times faster and takes $\approx 23.3 \text{ sec}$ compared to CS$_{64\times}$ with $\approx 17 \text{ min}$ for SM recovery. Still, some frequency components tend to have a high NRMSE (NRMSE $> 0.11$) for our 3d-SMRnet$_{64\times}$, whereas for the 3d-SMRnet$_{8\times}$ they do not. This problem can occur because of the equidistant subsampling and the symmetric patterns in the SM.

\begin{figure}[h!]
	   \vspace{-1em}
	   \centerline{\includegraphics[width=1\textwidth]{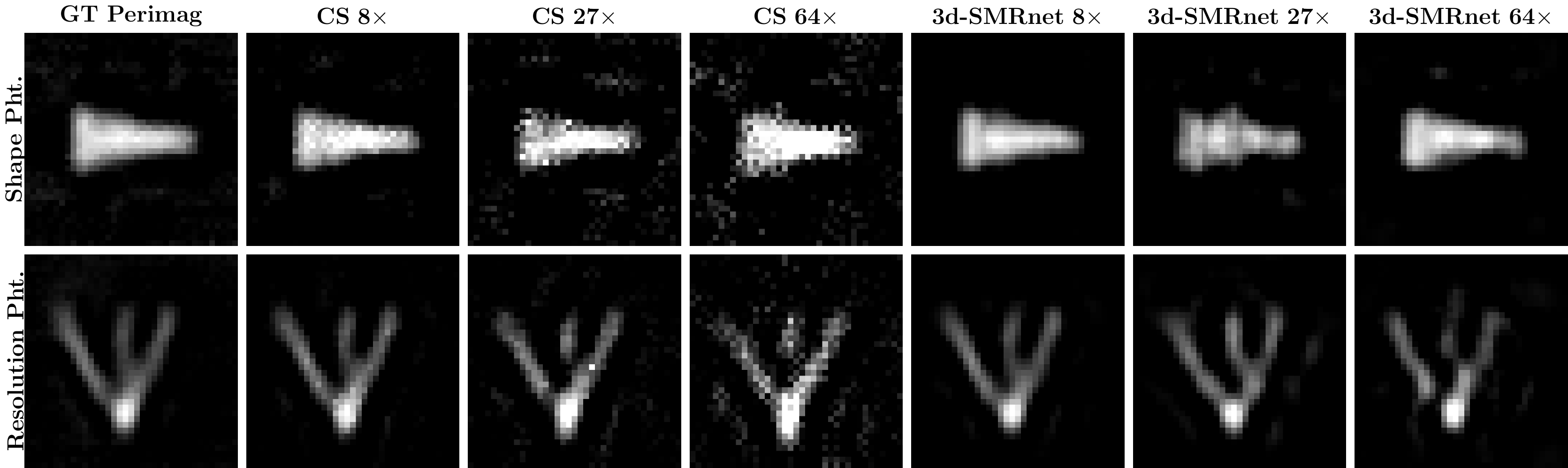}}
		\caption{Exemplary reconstruction of the measurements for the shape and resolution phantom with all recovered system matrices. We selected the slice $Z=19$.} 
		\label{fig:recoResults}
		\vspace{-2em}
	\end{figure}

     In Fig.~\ref{fig:recoResults}, we show one representative slice ($Z=19$) of the reconstructed shape and resolution phantoms. For CS, all reconstructed phantoms show a ''checkerboard`` noise and an overestimation of the voxel intensity, which increases for 27- and 64-fold subsampling. Our proposed 3d-SMRnet results produce smoother reconstructed images, with voxel intensities better resembling the ground truth data. Yet, the shape phantom reconstructed with the S$^{27\times}_{\text{3d-SMRnet}}$ shows some artifacts,  while for S$^{64\times}_{\text{3d-SMRnet}}$ those are not present. Still, the results for the resolution phantom with S$^{27\times}_{\text{3d-SMRnet}}$ are visually better than S$^{64\times}_{\text{3d-SMRnet}}$ (see Fig.~\ref{fig:recoResults} second row).
    Table~\ref{tab:imgMetricResults}. lists the subject-wise average structural similarity index (SSIM), peak signal to noise ratio (PSNR), and NRMSE. 
    For all three phantoms, the overall best results archived our 3d-SMRnet$_{8\times}$ with 0.0113, 0.9985, and 64.74 for $\overline{\text{NRMSE}}$, $\overline{\text{SSIM}}$, and $\overline{\text{PSNR}}$. Compared to the second best CS$_{8\times}$, this is an improvement by $31.1\%$ and $5.2\%$ for $\overline{\text{NRMSE}}$ and $\overline{\text{PSNR}}$. Furthermore, our 3d-SMRnet$_{64\times}$ is on par with CS$_{27\times}$ for the resolution phantom and considerably better for the shape and the concentration phantom.

    \begin{table}[h!]
        \vspace{-2em}
        
        \tiny
        \centering
        \setlength{\tabcolsep}{2pt}
        \begin{tabular}{l l l l | l l l | l l l  | l l l}
        & \multicolumn{3}{c}{\textbf{Shape Phantom}} & \multicolumn{3}{c}{\textbf{Resolution Phantom}} & \multicolumn{3}{c}{\textbf{Concentration Pht.}} & \multicolumn{3}{c}{\textbf{Avg.}}\\
	   & NRMSE & SSIM & PSNR & NRMSE & SSIM & PSNR & NRMSE & SSIM & PSNR & $\overline{\text{NRMSE}}$ & $\overline{\text{SSIM}}$ & $\overline{\text{PSNR}}$ \\
    	\hline
    	\hline
    	CS$_{8\times}$ & 0.0275 & 0.9908 & 51.64 & 0.0120 & 0.9996 & 67.00 & 0.0098 & 0.9995 & 65.93 & 0.0164 & 0.9966 & 61.52\\
     	CS$_{27\times}$ & 0.0628 & 0.9394 & 44.47 & 0.0267 & 0.9975 & 60.06& 0.0206 & 0.9972 & 59.42 &  0.0367 & 0.9780 & 54.60\\
     	CS$_{64\times}$ & 0.0915 & 0.8804 & 41.19 & 0.0452 & 0.9922 & 55.50& 0.0349 & 0.9916 & 54.86 &  0.0572 & 0.9547 & 50.52\\
     	3d-SMRnet$_{8\times}$ & 0.0186 & 0.9959 & 55.03& 0.0087 & 0.9998 & 69.79& 0.0066 & 0.9998 & 69.39 & 0.0113 & 0.9985 & 64.74\\
     	3d-SMRnet$_{27\times}$ & 0.0320 & 0.9866 & 50.31& 0.0208 & 0.9985 & 62.25& 0.0135 & 0.9988 & 63.09 & 0.0221 & 0.9946 & 58.55\\
     	3d-SMRnet$_{64\times}$ & 0.0284 & 0.9874 & 51.36& 0.0249 & 0.9978 & 60.68& 0.0150 & 0.9986 & 62.22 & 0.0228 & 0.9946 & 58.09
        \end{tabular} 
        \vspace*{0.2cm}
        \caption{Numerical results for all reconstructed phantoms demonstrates that our 3d-SMRnets clearly outperform CS when comparing the same subsampling factors. }
        \label{tab:imgMetricResults}
        \vspace{-2em}
    \end{table}
	
	\vspace{-1em}
	\section{Conclusion}
	We presented a novel method based on a 3d-System Matrix Recovery Network to significantly shorten calibration time in MPI. Our method can recover a highly subsampled system matrix: Using 64-times less samples compared to the original SM still allowed sufficient recovery of the SM. We have further shown that our method not only outperforms the current state-of-the-art in SM recovery quality (CS), but also in reconstructed image quality and processing time. Furthermore, our 3d-SMRnet can be applied to different types of particles after training. In the future, it is of interest to evaluate different kinds of sampling methods and multi-color MPI where two particles are simultaneously imaged.

	\bibliographystyle{splncs04}
	\bibliography{references}

\begin{thebibliography}{10}
\providecommand{\url}[1]{\texttt{#1}}
\providecommand{\urlprefix}{URL }
\providecommand{\doi}[1]{https://doi.org/#1}

\bibitem{donoho2006cs}
Cand\`{e}s, E., Romberg, J., Tao, T.: Robust uncertainty principles: Exact
  signal reconstruction from highly incomplete frequency information. IEEE
  Transactions on Information Theory  \textbf{52},  489--509 (2006)

\bibitem{Chen2018}
Chen, Y., Shi, F., Christodoulou, A.G., Xie, Y., Zhou, Z., Li, D.: Efficient
  and accurate mri super-resolution using a generative adversarial network and
  3d multi-level densely connected network. In: Medical Image Computing and
  Computer-Assisted Intervention. pp. 91--99. Springer International Publishing
  (2018)

\bibitem{Chen2018BrainMS}
Chen, Y., Xie, Y., Zhou, Z., Shi, F., Christodoulou, A.G., Li, D.: Brain mri
  super resolution using 3d deep densely connected neural networks.
  International Symposium on Biomedical Imaging pp. 739--742 (2018)

\bibitem{Dong2014}
Dong, C., Loy, C.C., He, K., Tang, X.: Learning a deep convolutional network
  for image super-resolution. In: European Conference on Computer Vision. pp.
  184--199. Springer International Publishing (2014)

\bibitem{ford1998colour}
Ford, A., Roberts, A.: Colour space conversions. Westminster University, London
   \textbf{1998},  1--31 (1998)

\bibitem{Gleich2005TomographicIU}
Gleich, B., Weizenecker, J.: Tomographic imaging using the nonlinear response
  of magnetic particles. Nature  \textbf{435},  1214--1217 (2005)

\bibitem{Knopp2010PhysMedBio}
Knopp, T., Rahmer, J., Sattel, T.F., Biederer, S., Weizenecker, J., Gleich, B.,
  Borgert, J., Buzug, T.M.: Weighted iterative reconstruction for magnetic
  particle imaging. Physics in Medicine and Biology  \textbf{55}(6),  1577 --
  1589 (2010)

\bibitem{knopp2013sparse}
Knopp, T., Weber, A.: Sparse reconstruction of the magnetic particle imaging
  system matrix. IEEE transactions on medical imaging  \textbf{32}(8),
  1473--1480 (2013)

\bibitem{ledig2017photo}
Ledig, C., Theis, L., Husz{\'a}r, F., Caballero, J., Cunningham, A., Acosta,
  A., Aitken, A., Tejani, A., Totz, J., Wang, Z., et~al.: Photo-realistic
  single image super-resolution using a generative adversarial network. In:
  Conference on Computer Vision and Pattern Recognition. pp. 4681--4690 (2017)

\bibitem{odena2016deconvolution}
Odena, A., Dumoulin, V., Olah, C.: Deconvolution and checkerboard artifacts.
  Distill  (2016)

\bibitem{Tai2017}
Tai, Y., Yang, J., Liu, X.: Image super-resolution via deep recursive residual
  network. In: Conference on Computer Vision and Pattern Recognition (2017)

\bibitem{Wang2016MRI}
{Wang}, S., {Su}, Z., {Ying}, L., {Peng}, X., {Zhu}, S., {Liang}, F., {Feng},
  D., {Liang}, D.: Accelerating magnetic resonance imaging via deep learning.
  In: International Symposium on Biomedical Imaging. pp. 514--517 (2016)

\bibitem{Wang2019ESRGAN}
Wang, X., Yu, K., Wu, S., Gu, J., Liu, Y., Dong, C., Qiao, Y., Loy, C.C.:
  Esrgan: Enhanced super-resolution generative adversarial networks. In:
  European Conference on Computer Vision Workshops. pp. 63--79. Springer
  International Publishing (2019)

\bibitem{weber2015reconstruction}
Weber, A., Knopp, T.: Reconstruction of the magnetic particle imaging system
  matrix using symmetries and compressed sensing. Advances in Mathematical
  Physics  (2015)

\end{thebibliography}
\end{document}